# The Hubbard Model:
# Exact Results in the Strong Coupling Limit

Donald M. Esterling[1]     VeritasCNC     Portland, OR 97209

## Abstract

Exact relations are derived for the Fermi Hubbard spectral weight function for infinite U at zero temperature in the thermodynamic limit for any dimension, any lattice structure and general hopping matrix. These relations involve moments of the spectral weight function but differ from similar work by (a) restricting the moments over the interesting low energy (lower Hubbard band) spectrum and  (b) without any of the usual approximations (e.g. decoupling) for the requisite higher order correlation functions.  The latter are determined exactly as explicit functions of the single particle correlation function in the infinite U limit.   The methods behind these results have the additional virtue of simplicity – tedious, but entirely straightforward.  In a companion paper, the relations offer rich insights into the model behavior and offer demanding quantitative and qualitative tests – a computer lab – of any proposed solution to the Hubbard model in the strong correlation (large finite U) region.  As the electron density approaches one electron per site, then the correlation functions become local, so are trivial to compute exactly.  In that limit and for the paramagnetic case, the second central moment of the spectral weight function is not zero, but quite large.

## Introduction

The (Fermi) Hubbard model [1] is a deceptively simple statement of a many body problem – becoming the object of theoretical treatments over many decades (e.g. as reviewed in [2-5]).  The model has been applied to various experimental phenomenon, but with conflicting conclusions about the underlying physics leading to "a great need for reasonably unbiased methods for determining the physics from a spectral analysis" [6].  Further, the single band Hubbard model (the most commonly treated case) is itself a simplification of materials with d-bands, p-band hybridization, random alloy effects and, of course, inter-site coulomb repulsion.  An "unbiased" arbiter of proposed model solutions, clean without conflating  how well the model itself replicates the experiment, would be of some value.

The interesting region for the Hubbard model is the strong coupling region (large intra-atomic coulomb repulsion U relative to the electron hopping  $\Delta$ – both U and $\Delta$ are defined below).  Weak interactions (small U) are a well-trodden area [7].  The challenge is to find a solution that is qualitatively and quantitatively correct in both the weak and the strong coupling region and then apply that solution to available experiments where the coupling may be intermediate in strength.

But a theory which proposes to be valid across weak to strong coupling as a minimum needs to be correct in the strong correlation limit.  Various theories have been proposed that claim to be exact in both the weak and strong coupling extremes, specifically those based on mean field theory [8-12]. We will demonstrate that this claim is not definitive, mostly due to a common misunderstanding of the "atomic limit" of the Hubbard model.

---

[1] Electronic address: don@veritascnc.com

Indeed, for some years, a number of authors have asserted that the Hubbard model [1] atomic limit is trivial and local, e.g. [10-12]. As has been established for decades [2, 13-14], the atomic limit is non-trivial, challenging and, most importantly for the current purpose, *non-local.* This has important implications for solutions away from the atomic limit.

Aside from the one dimensional case (e.g. [15-17]), there is a lack of exact benchmark results for the strong coupling region of the Hubbard model. Some that claim to be exact in fact then make approximations to the resulting expressions casting doubt on the results (e.g. Di Matteo and Claveau [18]. Exact benchmarks in the strong coupling limit would serve as valuable unbiased tests of the proposed theoretical solutions.

This paper offers just such exact relations with moments of the lower Hubbard band (LHB) spectral weight function (SWF), bereft of common approximations such as decoupling approaches [19-21]. The expressions offer rich insights into the strong coupling region. The exact moment expressions are derived as explicit functions of the equal time single particle correlation functions.

The exact results apply to the strong correlation limit (U goes to infinity, finite $\Delta$ where U is the on-site repulsion and $\Delta$ is the hopping as defined below) and zero temperature. But the results offer lessons for large finite U as well. Unless specified otherwise, the results apply to any dimension and any lattice structure. For convenience, we will limit electron densities to less than or equal to the number of sites. (Application to the upper Hubbard band is easily done with a similar technique). Extensions to finite temperature are straightforward, as long as the temperature ( $k_BT$ ) is small relative to the hopping $\Delta$ [7].

Since Hubbard introduced his eponymous model in 1963, there have been dozens if not hundreds of proposed solutions for the strong correlation region. LeBlanc, et al. 2015 [5] offer a convenient summary of many of the current solution methods. Many of the proposed solutions lack a rigorous foundation and so have an uncertain validity [2]. Over the last few years, the single site Dynamic Mean Field theory (DMFT) [8-11] and its extension to clusters [12 and references therein] have become a popular solution strategy. But a central assumption of DMFT is that the solution can be expressed with a local self energy. The cluster extensions allow a non-local contribution to the self energy, step-wise averaged over regions in momentum space. This paper and a companion paper [22] will offer a critique of the main DMFT assumption of a local self energy. The companion paper will combine the exact results in this paper with the strong coupling limit of DMFT and its cluster extensions to test the mean field theories in that limit. The results clearly indicate that DMFT and – quite possibly – cluster extensions, far from being exact, are problematic in the strong coupling region. Below and the companion paper [22] will also suggest that these mean field theories are not even exact in infinite dimensions as has been claimed [8-12].

These results follow closely the methods and results from an earlier paper by Esterling and Dubin [2] for the strong correlation region of the Hubbard model. The properties of the low energy spectrum (lower Hubbard band or LHB) of the spectral weight function (SWF) are analyzed in terms of moment expressions. In contrast to other moment approaches [19-21], the higher order correlation functions required for the moments are computed *exactly* (for infinite U) as functions of the single particle equal time correlation function. There is no decoupling or related assumptions, lacking in rigor.

It is well known that moments of a function alone cannot, in general, uniquely determine the explicit function. However the moments provide important insights and constraints on any solution in the strong correlation region. Further, as shown by Esterling [22], the exact moment relations can be used in two ways: [1] To generate a "best" single peak LHB SWF using only the sole assumption that there is only one main peak and [2] as a computer lab that offers an unbiased yet strong test of any proposed solution for the Hubbard model in the strong coupling region.

One of the more interesting results from this analysis is that, as the electron density approaches one electron per site (so the equal time correlation functions are local and trivial to compute), then for the paramagnetic case the second central moment of the SWF is not zero but is in fact quite large. Since the number of carriers (holes) approaches zero in this limit, experience in the weak coupling region suggests that the scattering should become small. If the width correlates with the lifetime or scattering, this is inconsistent with a non-zero width (second central moment) of a single peak LHB SWF. The resolution is to observe that no such correspondence is required in the strong coupling region. In fact, Esterling [22] observes there is even a large non-zero width (more precisely, second central moment) for the much simpler Falikov Kimball model [23] where the hopping is restricted to the up spin electrons only as the electron density approaches one electron per site.

The moment method is agnostic regarding the specific functional form of the SWF. An alternative form, consistent with our results, is a LHB SWF is described by two or more main peaks, not one main peak as in virtually all proposed solutions, aside from the one dimensional case. Some consideration of a multi-peak LHB SWF is given in the companion paper [22].

### Hubbard Atomic Limit

Since understanding the Hubbard atomic limit is central to our results, we offer a clarification of that term. Our definition of "atomic limit" is $\Delta / U \rightarrow 0$ at zero temperature.

There are 3 cases which fit this definition.

**AL0** – $\Delta$ identically zero, $k_B T > 0$ (but $\rightarrow 0$) and U finite. So $k_B T \gg \Delta$.

This is the "high temperature" limit ($k_B T \gg \Delta$). The commonly made statements that "the atomic limit is trivial (or local)" refer to this atomic limit. This limit has a factorial degeneracy (all of the ways to assign $N_e$ electrons to N sites, excluding doubly occupied sites). It is not correct to use this limit to validate solutions where the physically relevant atomic limits are **AL1** or **AL2** below. This (high temperature) atomic limit is of no further interest here.

**AL1** – $k_B T$ identically zero, $\Delta > 0$ (but $\to 0$) and U finite. So $\Delta \gg k_B T$.

This is the physically interesting "atomic limit". The ground state is unique[2], though typically not known. As shown here, correlation functions (such as $<c_{i\sigma}^\dagger c_{j\sigma}>$ and $<n_{i\sigma} n_{j\sigma}> - n_\sigma^2$) are not local.

**AL2** – $k_B T$ identically zero, $\Delta > 0$ and U infinite. So $\Delta \gg k_B T$.

This limit (finite $\Delta$, infinite U) is not usually referred to as an atomic limit. But the limit fits our definition. Our goal is to derive some rigorous results in the **AL2** region putting constraints on solutions for finite $\Delta$, finite U. The ground state is unique, though typically not known. Correlation functions are not local. **AL1** and **AL2** share some common features, but are not the same as noted below. However, as also noted below, equal time correlation functions for **AL1** and **AL2** are equal.

The physics behind the non-locality in the (**AL1** and **AL2**) atomic limit is easy to understand. For finite "hopping" $\Delta$ (defined below) by electrons between sites and for any intra-atomic electron repulsion U, the ground state is unique if not necessarily known. For exactly zero $\Delta$, there are an exponential number of degenerate ground states. For less than one electron per site, this consists of all of the ways that electrons can arrange themselves without two electrons on a single site. In the *atomic limit* (**AL1** or **AL2**), no matter how small $\Delta$ is relative to U, this degeneracy is broken and the system (e.g. various correlation functions) is non-local. The ground state is some linear combination of the degenerate ground states. An analogy with the free electron case may help. At zero temperature, for any $\Delta$, the momentum distribution is unity to the Fermi momentum and zero thereafter. The inter-site correlation function ($<c_{i\sigma}^\dagger c_{j\sigma}>$) is the Fourier transform of the momentum distribution, is not zero and has a RKKY-like form, dependent on the Fermi momentum (and so the electron density) but independent of $\Delta$. When U is not zero and for any finite $\Delta$, as with the free electron case, the momentum distribution is not constant but has some structure even for infinite U. This is shown explicitly in Fig. 2 of the companion paper [22]. Calculations for the 1D Hubbard model for infinite U [24-25] show a similar structure in the momentum distribution implying a similar non-locality in the atomic limit inter-site correlation function. This results in non-locality for single and multi-particle correlation functions and, as demonstrated below, an atomic limit ($\Delta \to 0$) that is not local and is not trivial, as assumed by many authors.

---

[2] Of course there is some degeneracy. Consider a free electron system. The same total energy can be obtained by selecting different $k_F$ points, each with the same Fermi energy. Here uniqueness simply means the absence of the factorial degeneracy of **AL0**.

## Solution Strategy

The genesis of this approach is the pioneering work by Harris and Lange [13] who derived rigorous expressions for the moments of each sub-peak (LHB and UHB) of the Hubbard frequency and momentum dependent spectral weight function (SWF) in the strong correlation region. Harris and Lange were one of the first to emphasize the non-trivial nature of the Hubbard atomic limit.

Using the Harris and Lange method, exact zeroth, first and second moments of the low energy peak in the spectral weight function for finite $\Delta$, infinite U are expressed in terms of equal time multi=particle correlation functions.

Because these are equal time correlation functions, there are no frequency dependent terms to consider which may lie hidden in and near the atomic limit. This is in contrast to the full time dependent Green's functions where the corresponding frequency dependence renders any perturbation solution of the Green's function as problematic [14,26-27][3]. The explicit $\Delta$'s that precede these multi-particle equal time correlation functions means that, in our infinite U limit, these correlation functions may be evaluated in the atomic limit.

As detailed in the next section and as we demonstrated some decades ago (E&D), while the correlation functions cannot be determined exactly in the atomic limit, we can generate exact expressions for equal time multi-particle correlation functions as functions of single particle equal time correlation functions.

But therein lies a problem. As already emphasized, the atomic limit is neither trivial nor local. How to evaluate these single particle correlation functions in that limit?

There are two alternatives.

The moment expressions can be used as a "computer lab." A proposed solution for the single particle SWF can be used to, first, directly compute moments of the LHB SWF (large U case) and, second, compute the required single particle equal time correlation functions then using these in the exact moment expressions to compute the moments. A comparison of the results offers substantial insights into the accuracy of the solution as shown in (Esterling 2018)

---

[3] As commented by Cyrot in his review [3], "…Many authors have… claimed that their procedures were exact to lowest order in $\Delta$. However Esterling ([14]) observes that any solution to the Hubbard model which is correct to lowest order in $\Delta$ would be equivalent to an exact solution of a certain excluded volume problem (Author: the infinite U model of this paper)… Esterling's observation strongly decreases the interest in this type of approach.." Nonetheless, authors continue to propose perturbation solutions in $\Delta$, e.g. by Metzner [28], though Metzner does observe that the "perturbation expansion…is only valid when the hopping matrix $t_{ij}$ is small compared to the temperature…" and Dai, Haule and Kotliar [29].

Another approach is to employ a self-consistent solution along the lines of (Potthoff and Nolting, 1996) but without their decoupling approximation.

The single particle Green's function can be evaluated from the SWF (Kadanoff and Baym, 1962). Once we adopt an explicit functional form (a single Gaussian in the companion paper [22] ) for the SWF, we have a closed set of self-consistent equations to solve. While not entirely trivial, the solution is straightforward and generates explicit numerical results for the general (frequency and momentum dependent) SWF and a self-consistent single particle Green's function as in [22].

This paper limits the discussion to the paramagnetic case of equal up and down electrons, though extensions to general spin densities is available. Also we will use the term "quasi-particle" to identify results different from the non-interacting electron case, not at all to infer that the quasi-particles will behave similarly to results for a weakly interacting system.

The zeroth order moment of the LHB SWF is simply (1-n) for all momentum, where n is the density of the up (or down) electrons.

The first order moment provides the "quasi-particle" energy (as a function of electron density and momentum). The quasi-particle energy can be derived from a wholly local self energy with only moderate non-local effects since the non-local component of this moment ($L_{ij}$ in [2], Eqn 12) is $O(n^2)$ or less based on ([2], Eqns. 25-27 ) while the leading terms are $O(n)$. As shown by Esterling [22], the local DMFT self energy is reasonably accurate for the first moment over all electron densities.

The second central moment can be related to the quasi-particle width (lifetime) or, more precisely, the imaginary part of the self energy [7]. As noted by Esterling [22], the DMFT SWF reduces to a single delta function in the strong coupling limit so the quasi-particle width is zero, contradicting the moment results. The self-consistent (single LHB SWF peak) numerical results and exact results (see below) as the electron density approaches one electron per site demonstrate that either the (single peak) SWF width is substantial or the LHB SWF must exhibit two or more main peaks in and near this limit.

## Technical Details

The model, formalism, notation and methods are the same as in Esterling and Dubin [2]. Virtually all of the technical details and notation are the same as in that paper, so they will not be replicated here. We do offer an alternative derivation of the atomic limit higher order Green's functions as functions of the single particle Green's function.

We seek the first and second central moments of the lower Hubbard peak as functions of the single particle equal time correlation function $< c_{i\sigma}^{\dagger} c_{j\sigma} >$ .

For reference, the Hubbard Hamiltonian is

$$H = \Delta \sum_{ij\sigma} t_{ij}\, c^\dagger_{i\sigma} c_{j\sigma} + \frac{1}{2} U \sum_{i\sigma} t_{ij}\, n_{i\sigma}\, n_{i\,-\sigma} \tag{1}$$

where $c^\dagger_{i\sigma}$ creates an electron on site i with spin σ, $c_{i\sigma}$ annihilates an electron on site i with spin σ, $n_{i\sigma}$ is the number operator for an electron on site i with spin σ, $t_{ij}$ is the usual hopping matrix, setting $t_{ii} = 0$ with no loss in generality (can be assumed into the Fermi energy), $\Delta$ scales the hopping term and U scales the intra-atomic repulsion. The hopping matrix $t_{ij}$ is completely general here, but the examples in the companion paper [22] will be for nearest neighbor coupling.

As can easily be verified by an expansion of the exponential

$$e^{iU\, n_{i\sigma}\, n_{i-\sigma}\, t} = n_{i\sigma}\, n_{i-\sigma}\, e^{iUt} + (1 - n_{i\sigma}\, n_{i-\sigma}) \tag{2}$$

Using this and again expanding the exponential in the following equation, in the atomic limit ($\Delta \to 0$) the explicit time dependence of the creation and annihilation operators become:

$$c_{i\sigma}(t) = e^{iHt} c_{i\sigma}(0) e^{-iHt} = (1 - n_{i-\sigma})\, c_{i\sigma}(0) + n_{i-\sigma}\, c_{i\sigma}(0)\, e^{-iUt} \tag{3a}$$

$$c^\dagger_{i\sigma}(t) = e^{iHt} c^\dagger_{i\sigma}(0) e^{-iHt} = (1 - n_{i-\sigma})\, c^\dagger_{i\sigma}(0) + n_{i-\sigma}\, c^\dagger_{i\sigma}(0)\, e^{+iUt} \tag{3b}$$

We are now prepared to demonstrate how higher order atomic limit correlation functions can be expressed as functions of the single particle correlation function. We do this for a particularly simple Green's function from Esterling and Dubin [2][4]. Reducing other higher order Green's functions in the **AL1** atomic limit follows a similar strategy but can become much more involved. In all cases in this paper, since we seek here the low energy peak in the Hubbard SWF, we will assume less than one electron per site and no doubly occupied sites in the ground state. A similar analysis can be done for the high energy SWF peak with more than one electron per site and no empty sites in the ground state.

Take $\Gamma_{ij\sigma} = -i <( n_{i-\sigma}\, c_{i\sigma}(t_i)\, c^\dagger_{j\sigma}(t_j) )_+ >$. Using the unequal time anticommutation relations from Equations ([2], 15a) and ([2], 15b) which follow directly from Equations (3a) and (3b) above, then if $R_i \neq R_j$ and noting that $n_{i-\sigma}\, c_{i\sigma}(t_i)\,|> = 0$ for no doubly occupied sites, $\Gamma_{ij\sigma}$ is zero unless $R_i = R_j$ and $t_i < t_j$. So $\Gamma_{ij\sigma} = -i\, \delta_{R_i\, R_j}\, \theta(t_i - t_j)\, <n_{i-\sigma}\, c_{i\sigma}(t_i)\, c^\dagger_{j\sigma}(t_j) >$. Inserting the expressions in Equations (3a) and (3b) and using $n_{i\sigma}\, n_{i-\sigma}\,|> = 0$ and $<|\, n_{i-\sigma}\,|> = n_{-\sigma}$ we end up with $\Gamma_{ij\sigma} = -i\, \delta_{R_i\, R_j}\, \theta(t_i - t_j)\, e^{-iU(t_i - t_j)}\, n_{-\sigma}$. This is a particularly simple case where the final result might be obtained more directly from an equation of motion. But we will require general two and three body Green's functions to express the related equal time correlation functions in the first and second moments of the SWF as functions of the single particle correlation function $<c_{i\sigma}^{\dagger}c_{j\sigma}>$. Those computations involve somewhat more complex manipulations, but still repeating the above process of reducing higher order Green's functions and equal time correlation functions to lower order functions. See for example Eqns. 19-27 in [2].

---

[4] Two errors should be noted in [2]. Eqn. (15a) in [2] is to be corrected for $R_1 = R_{1'}$. The correction follows immediately from Equations (3a) and (3b) above. That correction is not needed for this analysis. In addition, the line after Eqn. 22 in [2] refers to the "left-hand side" not the "right-hand side."

The next step is critical in the atomic limit calculation. Commonly, Green's functions may be determined by solving an equation of motion. But as explained by Esterling [14] and Esterling and Dubin [2], in the **AL1** atomic limit certain Green's functions contain a time independent component. Naively taking a time derivative to generate an equation of motion in the atomic limit can lose these time independent terms and lead to contradictions [14]. A proper way to handle general atomic limit Green's functions is encapsulated in going from the usual equation of motion Eqn. 8 in [2] to Eqn. 9 in [2]. The normal equation of motion is multiplied on both sides by the single particle Green's function. Integrating by parts, the equation of motion takes the form of Eqn. 9 in [2]. Note there is *no* explicit $\Delta$ in that equation. The equation is valid for any finite $\Delta$ and U and no contradictions due to time independent terms ensue. Both sides need to be evaluated carefully in the atomic limit. A pleasant feature is that, if the restriction of no doubly occupied sites is maintained, then the implicit integrals over time in Eqn. 9 of [2] – and its extension for third order Green's functions – for each of the "Γ" style functions in that equation remove all explicit dependence on U as well! This means we immediately have convenient expressions for the infinite U atomic limit.

The final step is to find explicit expressions for the LHB SWF moments in terms of higher order equal time correlation functions. While we could follow the prescriptions of Harris and Lange [13], their results simplify considerably for infinite U. In that limit, we may evaluate the moments by defining Hubbard operators as $X_{i\sigma} = (1 - n_{i-\sigma}) c_{i\sigma}$ and using the following Hamiltonian:

$$H = \Delta \sum_{ij\sigma} t_{ij} X_{i\sigma}^{\dagger} X_{j\sigma} \tag{4}$$

with the single particle Green's function

$$G_{ij\sigma} = -i <( X_{i\sigma}(t_i) X_{j\sigma}^{\dagger}(t_j) )_+ > \tag{5}$$

The first moment calculation is straightforward as presented in [2]. The second moment calculation is long and cumbersome. This paper relies on a second moment expression generated automatically using the very helpful Mathematica program DiracQ by Wright and Shastry [30] rather than the less reliable manual method used by Esterling and Dubin [2]. The DiracQ second moment result is in the Appendix[5].

One important comment: The SWF and associated atomic limit equal time correlation functions used in the moment expressions are for **AL2** (finite $\Delta$, infinite U). The explicit manipulations used to reduce the atomic limit multi-particle correlation functions to single particle correlation functions are for **AL1** (zero $\Delta$, finite U) as in Eqns. (3a-3b). The general (unequal time) Green's functions are not equal in these two limits as, for example, the time dependence of the Fermi operators is different. ("Time" offers a third (inverse) energy scale). But the moment relations require equal time correlation functions. For finite $\Delta$ and finite U but zero temperature, dimensional arguments require the equal time correlation functions to depend only on the ratio $\Delta/U$. There are only two energy scales so making one or the other large or small is simply a matter of changing energy units. The equal time functions are the same whether we take the limit $\Delta \to 0$ and finite U or U $\to \infty$, finite $\Delta$. That is the equal time correlation functions are the same whether $\Delta$ is, say, $10^{-6}$ and U = 1 or $\Delta$ = 1 and U $= 10^6$ since the final result depends only on the ratio of $\Delta/U$.

This concludes the "exact results" paper. A companion paper [22] will delve into the implications of these results. But it is useful to summarize some of the more significant *exact* results:

- The moments of the SWF can be expressed in terms of n-particle equal time atomic limit (**AL2**) correlation functions. The first three moments are evaluated explicitly in terms of n-particle equal time atomic limit correlation functions.

- Each of the n-particle equal time atomic limit (**AL2**) correlation functions can be expressed in terms of the 1-particle equal time atomic limit (**AL1**) correlation function.

Further, there are various exact relations among the equal time correlation functions. For example, from Eqns. 25-27 in [2], the difference between the parallel and anti-parallel density-density equal time correlation functions is equal to the spin-spin correlation functions or, in the notation of [2]

- $D_{12}^{\sigma,\sigma}$ - $D_{12}^{\sigma,-\sigma}$ = $S_{12,\sigma}$ even for finite $\Delta$, infinite U (**AL2**).

Finally a key result. For one electron per site, all of the equal time correlation functions are local (off-diagonal terms vanish) and are trivial. This allows us to compute exactly with *no* **AL1** manipulation of the equal time correlation functions – simply using the definition of the SWF (paramagnetic case, $n_\sigma$ = $n_{-\sigma}$ = $n$ (Z is the number of near neighbors)

---

[5] The Mathematica program along with other technical details including the somewhat complex explicit expressions for the moments in terms of the single particle equal time correlation functions may be obtained by contacting the Author.

- The width of the LHB SWF in this limit is $\Delta(Z\,n\,(2-n))^{1/2}$

Here "width" is defined as the *second central moment divided by the zeroth moment of the LHB SWF*, anticipating cases where the LHB SWF may consist of a skewed peak or more than one main peak. This perhaps surprising expression is simply baked into the Hubbard model itself. This exact result was first derived by Harris and Lange [13]) in their Eqn. (6.19)[6].

For comparison, the second central moment for the (Falikov and Kimball 1969) model – trivial to compute – for the paramagnetic case and one electron per site is $\Delta(Z\,n\,(1-n))^{1/2}$ as shown in [22].

The mean field theories claim to be exact in infinite dimensions. In order to make the computations converge, $\Delta$ is scaled as $Z^{-1/2}$ as the dimensions (and Z) grows. In infinite dimensions, the equal time correlation functions are again local and off-diagonal terms vanish. In this case, the exact LHB SWF width for general electron density (paramagnetic case) is simply $\Delta(Z\,n\,(2-n))^{1/2}$ [22]. This expression for the LHB SWF width does not go to zero as the dimension becomes large, in contradiction to the zero SWF width predicted by DMFT. However, the ratio of the LHB SWF width to the bandwidth ($\Delta *Z$) does go to zero as $Z^{-1/2}$.

## Conclusion

A general procedure to derive moment relations for the LHB SWF for infinite U, finite $\Delta$ and zero temperature for any dimension and any lattice structure is presented. The relations exactly express the moments as functions of the atomic limit single particle equal time correlation function. The current work is limited to the zeroth, first and second moments of the LHB SWF, but higher order moments (skewness, kurtosis,..) follow from the same procedure. A strength of the method is the simplicity. There are no assumptions hidden in obscure mathematics. The results follow directly from the definition of the SWF. The drawback is the computations while straightforward become algebraically onerous. The Mathematica program developed by Wright and Shastry [30] reduces the computations enormously. Nonetheless, the computation of third and higher order moments will be challenging.

It is well known that moments of a function alone cannot, in general, uniquely determine the explicit function. However SWF moments calculated according to the Technical Details section (as functions of the single particle equal time correlation functions) offer a rigorous test of any solution to the Hubbard model. The proposed single particle Green's function can be inserted into the exact SWF moment expressions. The resulting moments can compared with the moments calculated from an explicit integral of the proposed LHB SWF, weighted by the frequency to the appropriate power. Any difference between the two results provides a direct quantitative and qualitative test of the solution – an unbiased computer lab test of any proposed solution for the Hubbard model in the strong coupling region.

---

[6] The $\Delta$ in the Harris and Lange paper is equal to $\Delta\,Z^{1/2}$ in this paper.

**Acknowledgements**

With pleasure, we acknowledge the inspiration of R.V. Lange. Numerous helpful conversations with B. S. Shastry are also gratefully acknowledged.

## Appendix

The second moment expression derived using DiracQ by Wright and Shastry [30] is as follows, where $M2_{ii\sigma}$ is the site-diagonal second moment for an up spin electron and $M2_{ij\sigma}$ is the off-diagonal ($i \neq j$) second moment. The superscripts (e.g. [1]) indicate restrictions on the site indices in the expressions. Other than those site restrictions, there are implied sums over all repeated indices.

$$M2_{ii\sigma} = \Delta^2 * [\, t_{ip}\, t_{pi} *(1-n_{-\sigma}) + t_{im}\, t_{pi} <(n_{i\sigma} - n_{i-\sigma})\, c^\dagger_{p-\sigma}\, c_{m-\sigma}>^{[1]} - 2*\, t_{im}\, t_{pi} < c^\dagger_{i\sigma}\, c_{i-\sigma}\, c^\dagger_{p-\sigma}\, c_{m\sigma}>^{[1]}\,]$$

and for $i \neq j$ :

$$\begin{aligned}
M2_{ij\sigma} = \Delta^2 * [\, & t_{ip}\, t_{pj} * (1 - 3*n_{-\sigma}) + 2*\, t_{ij}^{\,2} < c^\dagger_{i-\sigma}\, c_{j-\sigma}> - 2*\, t_{ij}\, t_{im} < c^\dagger_{i-\sigma}\, c_{m-\sigma}>^{[2]} \\
& + 2*\, t_{ij}\, t_{im} (< c^\dagger_{i-\sigma}\, c_{m-\sigma}\, n_{j-\sigma}>^{[2]} + < c^\dagger_{i-\sigma}\, c_{m\sigma}\, c^\dagger_{j\sigma}\, c_{j-\sigma}>^{[2]}\,) \\
& + t_{ij}\, t_{pj} * (< n_{i-\sigma}\, (c^\dagger_{j-\sigma}\, c_{p-\sigma} - c^\dagger_{p-\sigma}\, c_{j-\sigma})>^{[4]} + < c^\dagger_{i-\sigma}\, c_{i\sigma}\, (c^\dagger_{j\sigma}\, c_{p-\sigma} - c^\dagger_{p\sigma}\, c_{j-\sigma})>^{[4]}\,) \\
& + t_{ip}\, t_{pj} * (< n_{i-\sigma}\, n_{j-\sigma}> + < n_{i-\sigma}\, n_{p-\sigma}>^{[3]} + < n_{p-\sigma}\, n_{j-\sigma}>^{[4]}\,) \\
& + t_{ip}\, t_{pj} * (< c^\dagger_{i-\sigma}\, c_{i\sigma}\, c^\dagger_{j\sigma}\, c_{j-\sigma}> + < c^\dagger_{i-\sigma}\, c_{i\sigma}\, c^\dagger_{p\sigma}\, c_{p-\sigma}>^{[3]} + < c^\dagger_{p-\sigma}\, c_{p\sigma}\, c^\dagger_{j\sigma}\, c_{j-\sigma}>^{[4]}\,) \\
& - t_{ip}\, t_{pj} < n_{i-\sigma}\, n_{p-\sigma}\, n_{j-\sigma}> \\
& - t_{ip}\, t_{pj} (< c^\dagger_{i-\sigma}\, c_{i\sigma}\, c^\dagger_{p\sigma}\, c_{p-\sigma}\, n_{j-\sigma}> + < c^\dagger_{i-\sigma}\, c_{i\sigma}\, c^\dagger_{j\sigma}\, c_{j-\sigma}\, n_{p-\sigma}> + < c^\dagger_{p-\sigma}\, c_{p\sigma}\, c^\dagger_{j\sigma}\, c_{j-\sigma}\, n_{i-\sigma}>\,)\,]
\end{aligned}$$

[1]   $m \neq p$       [2]   $m \neq j$       [3]   $i \neq p$       [4]   $j \neq p$

For reference and for use in the companion paper [22] for the case of the electron density approaching one electron per site, the inter-site correlation functions are all set to zero (e.g. $< c^\dagger_{i-\sigma}\, c_{j-\sigma}> = 0$  and $< n_{i-\sigma}\, n_{j-\sigma}> = n_{-\sigma}^2$ for $i \neq j$) and for general $i$ and $j$ and using the zeroth LHB SWF moment $M0_\sigma = (1-n_{-\sigma})$

$$\begin{aligned}
M2_{ij\sigma} / M0_\sigma &= \Delta^2 * [t_{ip}\, t_{pj} * (1 - n_{-\sigma})^2 + \delta_{ij} * t_{ip}\, t_{pi} * (1 - (1 - n_{-\sigma})^2\,)] \\
&= \Delta^2 * t_{ip}\, t_{pj} * [(1 - n_{-\sigma})^2 + \delta_{ij} * n_{-\sigma} *(2 - n_{-\sigma})]
\end{aligned}$$